\documentclass[twocolumn]{aastex631}

\shorttitle{Electron-Ion Equilibration in SNR Shocks}
\shortauthors{authors}
\graphicspath{{./}{figures/}}

\begin{document}

\def\RSUN{R$_{\sun}$ }
\def\kms{$\rm km~s^{-1}$}
\def\cmcube{$\rm cm^{-3}$}

\def\teti{T$_e$/T$_p$}
\def\ibin{I$_B$/I$_N$}
\def\Ha{H$\alpha$}

\title{Electron-Ion Temperature Ratio in Astrophysical Shocks}

\correspondingauthor{John C. Raymond}
\email{jraymond@cfa.harvard.edu}

\author[0000-0002-7868-1622]{John C. Raymond}
\affiliation{Center for Astrophysics $|$ Harvard \& Smithsonian,
60 Garden St.,
Cambridge, MA 02138, USA}

\author[0000-0002-9886-0839]{Parviz Ghavamian}
\affiliation{Towson University,
8000 York Road,
Towson, MD 221252, USA}

\author[0000-0002-5680-0766]{Artem Bohdan}
\affiliation{Max-Planck-Institut für Plasmaphysik, 
Boltzmannstr. 2, DE-85748 Garching, Germany}

\author[0000-0002-5455-2957]{Dongsu Ryu}
\affiliation{Department of Physics, College of Natural Sciences, UNIST, Ulsan 44919, Korea}

\author[0000-0001-6036-8569]{Jacek Niemiec}
\affiliation{Institute of Nuclear Physics Polish Academy of Sciences, PL-31342 Krakow, Poland}

\author[0000-0002-1227-2754]{Lorenzo Sironi}
\affiliation{Department of Astronomy, Columbia University, 550 W 120th St. MC 5246, New York, NY 10027, USA}

\author[0000-0003-3483-4890]{Aaron Tran}
\affiliation{Department of Astronomy, Columbia University, 550 W 120th St. MC 5246, New York, NY 10027, USA}

\author[0000-0002-9881-8112]{Elena Amato}
\affiliation{INAF-Osservatorio Astrofisico di Arcetri, Largo E. Fermi 5, 50125 Firenze, Italy}

\author[0000-0002-1818-9927]{Masahiro Hoshino}
\affiliation{Department of Earth and Planetary Science, The University of Tokyo, Tokyo 113-0033, Japan}

\author[0000-0001-7861-1707]{Martin Pohl}
\affiliation{University of Potsdam, Institute of Physics and Astronomy, D-14476 Potsdam, Germany}
\affiliation{DESY, D-15738 Zeuthen, Germany}

\author[0000-0002-2140-6961]{Takanobu Amano}
\affiliation{Department of Earth and Planetary Science, The University of Tokyo, Tokyo, 113-0033, Japan}

\author[0000-0002-8502-5535]{Federico Fiuza}
\affiliation{High Energy Density Science Division, SLAC National Accelerator Laboratory, Menlo Park, CA, USA}



\begin{abstract}
Collisionless shock waves in supernova remnants and the solar wind heat electrons less effectively than they heat ions, as is predicted by kinetic simulations.  However, the values of \teti\/ inferred from the \Ha\/ profiles of supernova remnant shocks behave differently as a function of Mach number or Alfv\'{e}n Mach number than what is measured in the solar wind or predicted by simulations.  Here we determine \teti\/ for supernova remnant shocks using \Ha\/ profiles, shock speeds from proper motions, and electron temperatures from X-ray spectra.  We also improve the estimates of sound speed and Alfv\'{e}n speed used to determine Mach numbers.  We find that the \Ha\/ determinations are robust and that the discrepancies among supernova remnant shocks, solar wind shocks and computer-simulated shocks remain.  We discuss some possible contributing factors, including shock precursors, turbulence and varying preshock conditions.

\end{abstract}




\section{Introduction} \label{sec:intro}

Collisional shock waves are characterized by a thickness of order the mean free path. Because the shock transitions are mediated by particle-particle collisions, they produce Maxwellian distributions and thermal equilibrium among the particle species.  On the other hand, the thickness of a collisionless shock in a plasma is much smaller, of order the proton gyroradius or ion skin depth.  Because they are mediated by electromagnetic fields and plasma turbulence instead of particle collisions, collisionless shocks can produce non-Maxwellian velocity distributions and very different temperatures among different particle species.  Except for shocks inside stars or in molecular clouds, most astrophysical shocks are collisionless. 

\begin{figure*}
\begin{centering}
\includegraphics[width=5.0in]{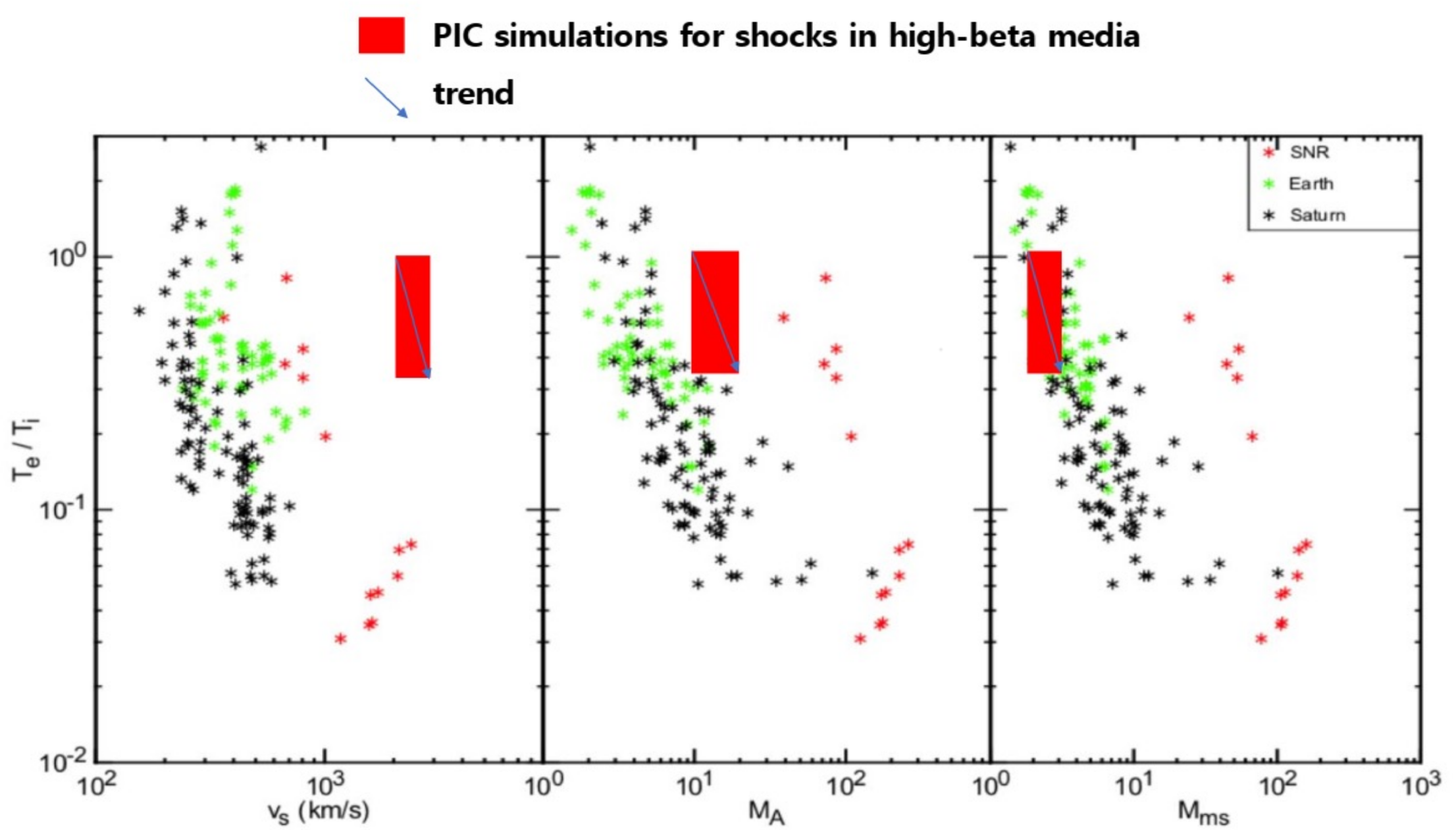}
\caption{Comparison of estimates of \teti\/ from SNR shocks and from solar system shocks measured in situ plotted against shock speed, Alfv\'{e}n Mach number and magnetosonic Mach number, adapted from Figure 7 of \citet{ghavamian13}.  The red boxes and arrows indicate trends from PIC simulations of low Mach number perpendicular shocks in the high $\beta$ plasma typical of galaxy clusters (Ryu et al., in preparation).  We expect $\beta$ in the range 0.2 to 2 for most SNR shocks, but the simulation values of \teti\/ show little dependence on $\beta$.
\label{ghavamian}
}
\end{centering}

\end{figure*}
The electron and proton temperatures T$_e$ and T$_p$ predicted by the Rankine-Hugoniot jump conditions can be very different, because a shock thermalizes most of the kinetic energy of the particles flowing through it, and the electron and ion kinetic energies differ by their mass ratio.  Collisions can eventually bring T$_e$ and T$_p$ into equilibrium, but over a time scale that can exceed the dynamical age of the astrophysical object (e.g., supernova remnants, galaxy cluster shocks, structure formation shocks).  However, energy dissipation in the shock front and precursor occurs by way of strong plasma turbulence, and this plasma turbulence is capable of transferring energy between ions and electrons more rapidly.

Reliable determination of \teti\/ is important for understanding the physics of collisionless shocks.  For instance, electron-ion thermal equilibration can affect the shock reformation process \citep{shimada05}. \teti\/ is also important as a diagnostic tool, since T$_e$ is measured from X-ray spectra. In partially neutral preshock conditions, shock speeds can also be determined from the \Ha\/ line profile, which effectively measures T$_p$ \citep{chevalier78, bychkov79}.  In cases where protons keep all the thermal energy, T$_p$ is twice as large as it will be in thermal equilibrium, when half the thermal energy is given to the electrons.  Finally, electron heating may play a significant role in the injection of particles into the diffusive shock acceleration process (DSA) which produces radio and X-ray synchrotron emission as well as gamma-ray emission from inverse Compton scattering in SNRs. 

Different estimates for the degree of electron-ion temperature equilibration in collisionless shocks give contradictory results.  The in situ measurements of bow shocks in the solar wind at Earth and Saturn show \teti\/ dropping from 1 at very low sonic Mach numbers or Alfv\'{e}n Mach numbers (M or M$_A$) to about 0.1 at Mach numbers near 10.  Although observations of supernova remnants (SNRs) show a parallel decline, the trend is systematically shifted to much higher Mach numbers.  In SNRs, \teti \/ is near 1 at M$\sim$30, and it drops to~0.05 at M$\sim$300 as shown in Figure~\ref{ghavamian} (adapted from \citet{ghavamian13}).

A second inconsistency arises at higher Mach numbers.  Both particle-in-cell (PIC) simulations and solar wind shocks show a drop in \teti\/  with Mach number between M=1 and M=15, but then \teti\/  rises to about 0.2-0.3 \citep{bohdan20, hanusch20, tsiolis21, wilson20}.  On the other hand, values of \teti\/ from the \Ha\/ profiles observed in SNR shocks continue to decline with M to about 0.05 at M$\simeq$300 \citep{ghavamian13}.
  
Given the importance of understanding the electron-ion temperature ratio for interpreting observations of SNRs and other shocks, as well as for understanding collisionless shock physics and the acceleration of nonthermal electrons in shock waves, we try to resolve these discrepancies.  In principle, the values of \teti\/ derived from the SNR observations, PIC simulations and in situ measurements could all be correct, but the former pertain to a structure many orders of magnitude thicker than do the latter.  The SNR measurements could be appropriate for such tasks as interpreting X-ray spectra, while the in situ and PIC results could be appropriate for studies of plasma processes on small scales.  For instance, the PIC simulations and solar wind measurements could pertain to a subshock, while the SNR observations would include any precursor or postcursor effects.   

On the other hand, the discrepancies could result from errors in either the theory or observations of the SNRs.  In particular, the comparison in \citet{ghavamian13} relied on the \Ha\/ line profiles of shocks in partially neutral gas.  The profiles consist of a narrow component from H atoms that are collisionally excited after they pass through the shock, and a broad component from atoms that undergo charge transfer with postshock protons, acquiring a velocity distribution like that of the protons before they are excited \citep{chevalier80}.  The ratio of the broad and narrow component intensities, \ibin, depends on the electron temperature, and it is often used to determine \teti .  However, this carries some uncertainty.

The aim of this paper is to compare \teti\/ determinations based on the intensity ratios of the broad and narrow components of the \Ha\/ profiles of SNR shocks \citep{ghavamian13} with determinations that use electron temperatures measured from X-ray spectra, T$_{e,x}$, proton temperatures based on H$\alpha$ broad component widths, T$_{p,w}$, and shock speeds based upon proper motions for SNRs at known distances, V$_{pm}$.  We make four sets of determinations based on 1) \ibin \,vs broad component \Ha\/ width, 2) T$_{p,w}$ vs T$_{e,x}$, 3)  V$_{pm}$ vs T$_{p,w}$, and 4) V$_{pm}$ vs. T$_{e,x}$, all based on published measurements of the relevant quantities.  We will look for common trends of \teti\/ with shock speed M or M$_A$ among these different estimates.  

\begin{figure*}
\begin{centering}
\includegraphics[width=5.0in]{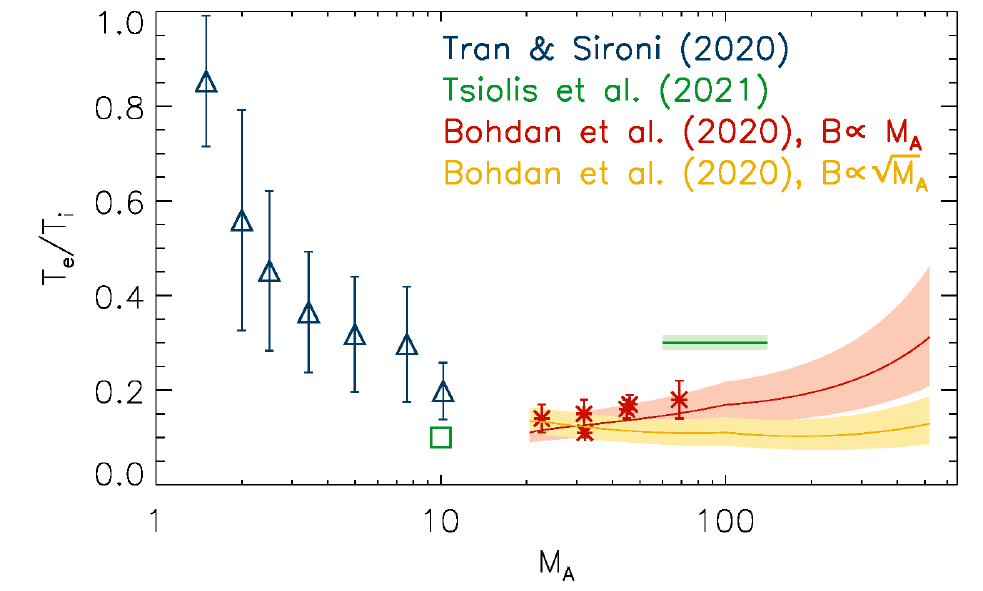}
\caption{\teti\/ for perpendicular shocks based on PIC \citet{tran20} (blue), \citet{bohdan20} (red) and \citet{tsiolis21} (green).  The prediction of Te/Tp at M$_A >$100 depends on the magnetic field generated by the Weibel instability: The red line is for B $\sim$ M$_A$ and the yellow line is for B $\sim$ $M_A^{1/2} $ (see updated results in \citet{bohdan21}).
\label{bohdan}
}
\end{centering}
\end{figure*}

We will not attempt a statistical comparison because the uncertainties are dominated by systematic errors in the measurements or the \teti\/ determinations that we cannot quantify.  For example, we try to choose a portion of an SNR shock where V$_{pm}$ and T$_e$ are both measured, but the values will not always pertain to exactly the same plasma.  All the values of the basic parameters are taken from the literature, and when a range of values is given, it indicates the range that entered an average rather than a 1-$\sigma$ or 3-$\sigma$ uncertainty.  The comparison is complicated by the fact that we often take, for instance, proper motions from one paper and electron temperatures based on X-ray spectra from another.  Detailed considerations for each SNR are given in the Appendix.  

The paper is organized as follows.  Section 2 describes predictions from numerical simulations, and Section 3 describes results from in situ measurements of shocks in the solar wind and in the laboratory.  Section 4 describes the diagnostic measurements of SNR shocks, and section 5 discusses the four methods of estimating \teti\/ in SNR shocks, along with the uncertainties involved.  Section 6 discusses the comparisons among SNRs, solar wind and laboratory shocks and theory.  It emphasizes the questions of whether the inclusion of shock precursors or postshock processes, cosmic ray acceleration, or the presence of neutrals in the regions observed in the SNRs can account for the differences.  Section 7 summarizes our results, and the Appendix gives details of the \teti\/ estimates for each SNR.

\section{Predictions from Theory and Simulations}

Early theoretical calculations for solar wind shocks based on energy transfer by various wave modes or on shock electric fields predicted \teti$\sim$0.2 \citep{cargill88, hull00}.  On the other hand, \citet{ghavamian07b} attempted to explain the observed decline on \teti\/ in SNRs with shock speed based on lower hybrid waves in a cosmic ray-generated precursor.

PIC simulations of shocks have explored part of parameter space.  For perpendicular shocks in plasma with $\beta$ near 1, \teti\/ drops steeply from 1 at very small M$_A$ to 0.1-0.2 at M$_A$ = 10 \citep{tran20}, then increases slowly from $\sim$0.1 at M$_A$ = 20 to 0.2 at M$_A$ = 100 \citep{bohdan20}, and perhaps tends asymptotically to 0.3 at higher M$_A$ \citep{tsiolis21}. It should be noted that the prediction for Mach numbers above 100 depends on the scaling with M$_A$, and the saturation level may drop from 0.3 to 0.1. The results are summarized in Figure 2.

Oblique shocks seem to behave like perpendicular shocks, even in the presence of the electron foreshock generated by the shock reflected electrons \citep{morris23}.  \citet{lezhnin21} report \teti$\sim$0.5 in a Mach 15 shock at $\Theta_{BN}=60^\circ$, where $\Theta_{BN}$ is the angle between shock normal and the magnetic field. On the other hand, parallel shocks are analytically predicted to show a decline in \teti\/ with Mach number similar to that seen in SNRs, as shown in Figure 12 of \citet{arbutina21}, or to show a decline from near equilibration to around 0.3 at M$_A$ around 20 \citep{hanusch20}.  \citet{tran20} investigated low M shocks in the context of the solar wind.  That work and subsequent simulations show that at low Mach numbers, the electron heating is not much above that predicted for adiabatic compression, but that the non-adiabatic heating decreases with increasing magnetic obliquity.  It increases with Mach number, and neither plasma $\beta$ nor the assumed electron-ion mass ratio is very important. 

PIC simulations have also been performed for weak shocks ($M\sim 2-5$) in high beta plasmas ($\beta\gg 1$), as appropriate to shocks that form in hot intracluster medium (ICM) during mergers of galaxy clusters. Electron heating mechanisms and their efficiency were investigated for strictly perpendicular 2D shocks by \citet{guo2017,guo2018} who noted that \teti\/ is independent of plasma $\beta$ in the range $\beta=4-32$ but decreases strongly with Mach number from \teti$\sim 0.8$ at $M=2$ to \teti$\sim 0.25$ at $M=5$. This result should hold at oblique superluminal shocks with $\Theta_{BN}>\Theta_{BN,cr} = \cos^{-1}{(v_{\mathrm{sh}}^{\mathrm{up}}/c)}$. 
A recent calculation by \citet{Kobzar2021} for an $M=3$ ($M_A$=6.1) and $\beta=5$ shock with a subluminal obliquity $\Theta_{BN}=75^{\circ}$, close to $\Theta_{BN,cr}\approx 81.4^{\circ}$ for their parameters, gives \teti$\sim 0.4-0.45$ in agreement with results by \citet{guo2018}.  This is also compatible with 2D PIC simulations presented in Kang et al. (2019a). They found (private communication) that \teti\/ increases with plasma beta from \teti$\sim 0.55$ at $\beta=20$ to \teti$\gtrsim 0.7$ at $\beta=100$ at a subluminal angle $\Theta_{BN}=63^{\circ}$ for an $M=3$ shock. \teti\/ decreases in their simulations with the Mach number, but can reach \teti$\gtrsim 1$ in very weak shocks with $M\sim 2$. It also depends quite strongly on $\Theta_{BN}$, and approaches \teti\/ around 1 at high obliquity angles in very high $\beta$ plasmas. 

A limitation of some of the PIC simulations is an unrealistic electron-ion mass ratio.  For instance, \citet{shalaby22} simulated very fast shocks (40,000 \kms) and found that a mass ratio of 100 suppresses an intermediate-scale instability in low M$_A$ shocks.  On the other hand, simulations by \citet{tran20} (m$_i$/m$_e$=20-625) and \citet{bohdan20} (m$_i$/m$_e$=50-400) demonstrate that the ion-to-electron mass ratio has a minor influence on \teti\/ if the mass ratio is high enough to separate electron and ion scales (e.g., m$_i$/m$_e>$100). They indicate that changing shock velocity ratio with respect to c, with Mach
and $\beta$ fixed, does not affect this conclusion \citep{lezhnin21}

They indicate that shock velocity does not affect this conclusion as long as all plasma components remain nonrelativistic.

\section{Measurements of Shocks in the Solar Wind and the Laboratory}

Extensive studies of \teti\/ in the Earth's bow shock \citep{schwartz88} and Saturn's bow shock \citep{masters11} have been conducted, and results are presented in \citet{ghavamian13} and in Figure~\ref{ghavamian}.  The recent work of \citet{wilson20} is based on careful analysis of 15,000 velocity distribution functions in 52 interplanetary shocks seen by the Wind spacecraft.  While there is a very large scatter, the average shows the electron heating increasing gradually with M$_A$, such that for solar wind shocks, \teti\/ would be expected to drop from 1 at small Mach numbers to $\sim$0.05 at Mach numbers near 10, then rise to around 0.4 at Mach numbers of order 40.  They found little correlation with parameters such as pre-shock $\beta$ or shock obliquity.  

As with the values of \teti\/ inferred from SNR observations and the values expected from PIC simulations, there are caveats.  In particular, the shocks may not be steady, and the electrons may be mobile enough that they reflect average conditions rather than the local shock parameters, so that measurements from a single spacecraft passing through a shock will not always reflect the expected local electron-ion equilibration.

Recent developments in high-power lasers are now creating important opportunities to probe the plasma microphysics of collisionless shocks in controlled laboratory experiments. The interest in using laser-produced plasmas to study the physics of collisionless shocks is not new. Early experiments in the 70's and 80's have used 100 J class lasers to ablate solid targets and produce interpenetrating plasma flows to study their collisionless coupling \citep{dean71, cheung73, bell88}. However, for such laser energies, these earlier studies, were limited to relatively low Mach numbers and electrostatic coupling. The development of high-energy laser facilities capable of delivering 10 kJ to MJ laser energy on target, such as OMEGA and the National Ignition Facility (NIF) are transforming the ability to study for the first time high-Mach number collisionless regimes dominated by electromagnetic processes as pertinent to most space and astrophysical shocks.

In the last few years, several experiments produced supersonic, super-Alfvénic  
plasma flows in the laboratory to study the underlying shock formation processes \citep{ross12, fox13, huntington15, niemann14, schaeffer17, rigby18, swadling20, fiuza20}. The plasma flow velocities produced are typically in the range of 500 \kms\/ to 2000 \kms\/ and the corresponding Mach number M = 2 - 400. Developments in plasma diagnostics are starting to allow resolved measurements of the evolution of the plasma temperature in these systems. In particular, recent experiments at NIF characterized the formation of a collisionless shock with M $\sim$ 12 and M$_A ~\sim$ 400, measuring a downstream T$_e$ = 3 keV using Thomson scattering, and demonstrating the acceleration of nonthermal electrons to energy $>$ 100 T$_e$ \citep{fiuza20}.  While these experiments did not measure T$_i$, considering that during shock formation the plasma flow velocity varies between 1800 km/s and 1000 km/s (the velocity of the laser produced flows varies in time) and using the standard conservation equations at the shock, one obtains $\rm Z T_e/T_i = 0.15-0.75$ (note that the experiments use carbon plasmas with Z = 6). This is consistent with the PIC results summarized in Fig. 2, which suggest T$_e$/T$_p$ = 0.1-0.4 for M$_A$ = 400. Planned follow-up experiments will measure simultaneously the electron plasma wave and ion acoustic wave features of the Thomson scattering spectrum \citep{ross12}, enabling a detailed characterization of T$_e$/T$_i$. Furthermore, by varying the field strength and orientation of an external magnetic field produced by coils it will be possible to study T$_e$/T$_i$ as function of M, M$_A$, and $\theta_{BN}$.

\begin{table*}
\centerline{Table 1}
\centerline{Equilibration from the H$\alpha$ \ibin }
%
\vspace*{0.1in}
\begin{centering}
\begin{tabular}{l c c c c c c c l}
\hline\hline
SNR & Pos & N$_{avg}$ & FWHM$_B$ & FWHM$_N$ & $\rm I_B/I_N$ & V$_{S}$ & \teti  & Ref \\
   &  &  & \kms & \kms &  & \kms & \\
\hline
\\
SN1006 & NW & 19 & 2300-2600 & 18-24 & 0.67-0.86 & 2700-3400 & 0.09 & N13, M12 \\ 
Tycho  & knot "g" & 4 & 1760-2100 & 49(250) &  0.67-1.08 & 1850-2400 & 0.05-0.15  & G01, K17, S91, M12   \\ 
Kepler & - & 6 & 1540-3500 & 42 & 0.45-1.1 & 1250-2700 & 0.8 & S03, B91, M12      \\
RCW 86 & SW & 1 & 544-580 & 36 & 1.15-1.21 &  580-620 & 0.35-0.65  & G01, S03             \\
Cygnus Loop &   North  &  9 & 203-296 & 35-43 & 0.60-1.39 &  300 & 0.8-1.0 & M14, G99 \\ 
0509-67.5     & NE & 3 & 4750-4235 & 25-31 & 0.05-0.7 & $>$7000 & $<$0.25 & H18, M12   \\
0519-69.0     & - & 1 & 1800-2800 & 39-42 & 0.4-0.8 & 2500 & $<$0.20 & S91, S94, M12, H18    \\
0548-70.4     & - & 1 & 720-800 & 39-58 & 0.9-1.3 & 760 & $>$0.5 & S91, S94, M12    \\
DEM L71  & - & 21 & 450-950 & 32-43 & 0.37-0.68 & 450-950 & $>$0.1 & S94, R09, M12   \\
1E102.2-7219    &  & & & & & & &  \\

\\
\hline
\vspace*{0.1in}

\end{tabular}

References for Tables 1-5:  
A22- \citet{alan22}, B19-\citet{bandiera19}, B91-\citet{blair91}, B06-\citet{borkowski06},
C22-\citet{coffin22}, G99-\citet{ghavamian99}, G00-\citet{ghavamian00}, 
G01-\citet{ghavamian01}, G17-\citet{ghavamian17}, G22-\citet{guest22a}, H02-\citet{hwang02}, 
H11-\citet{helder11}, H18-\citet{hovey18}, 
K78-\citet{kamper78}, K15- \citet{katsuda15}, K16- \citet{katsuda16}, K17-\citet{knezevic17}, K21-\citet{knezevic21},  K10-\citet{kosenko10}, 
Mi12-\citet{miceli12}, M12-\citet{morlino12}, M13-\citet{morlino13}, 
M14-\citet{medina14},
N13-\citet{nikolic13}, R03-\citet{rakowski03}, R09-\citet{rakowski09}, R17-\citet{raymond17}, S09-\citet{salvesen09}, S17 - \citet{sano17}, S16-\citet{schenck16},
S21-\citet{seitenzahl21}, S94-\citet{smith94}, S91 - \citet{smith91}, S03-\citet{sollerman03}, W13-\citet{winkler13},
W16-\citet{williams16}, W22-\citet{williams22}, W03-\citet{winkler03}, 
W13-\citet{winkler13}, X19-\citet{xi19}

\end{centering}
\end{table*}

\section{Diagnostic measurements}

Here we list the 3 diagnostic measurements we will be using.  Each measurement has systematic uncertainties.  We discuss some of them here, and we will discuss others in more detail in the next section in the context of the specific estimates of \teti.

Some diagnostics for determining \teti\/ in SNR shocks rely on the \Ha\/ profiles.  A collisionless shock in partially neutral gas produces a thin filament of emission in \Ha\/ and other hydrogen lines, because some of the neutral H atoms are excited before they are ionized \citep{chevalier78, chevalier80}.  The collisionless shock thickness is set by the proton gyroradius or the skin depth, and it is very thin: $\sim 10^{8-10} \rm ~cm$ for typical ISM magnetic fields and densities.  For comparison, the length scale for excitation and ionization of H is of order $10^{15} \rm ~cm$. The neutrals pass through the collisionless shock without feeling the electromagnetic fields or plasma turbulence, so many of them still have their preshock velocities when they are excited and emit \Ha.  On the other hand, some of the neutrals undergo charge transfer reactions with postshock protons.  Those neutrals acquire a velocity distribution similar to the proton distribution, though weighted by the charge transfer cross section and the relative velocity \citep{chevalier80, ghavamian01, heng07, vanadelsberg08, morlino13}.  Thus the \Ha\/ profile consists of a broad component, whose FWHM is comparable to the shock speed, and a narrow component, whose FWHM is typically 20-50 \kms\, \citep{sollerman03} .  

The width of the \Ha\/ broad component is a direct measurement of T$_p$ in the postshock region, and the narrow component width indicates the temperature in the preshock region.  Both widths include any turbulent motion.  The turbulence is likely to be significant in a shock precursor (e.g., \citet{bell2004}).  There may be significant turbulence in the postshock region as well.   Small scale (kinetic scale) turbulence will decay over a distance much smaller than the size of the \Ha\/ emitting region, but we discuss turbulence in more detail below.  The intensity ratio \ibin\/ depends on the ratio of ionization time to charge transfer time scale, and it therefore depends on the electron temperature \citep{raymond91, smith91, laming96, ghavamian01}.  


A second diagnostic is measurement of T$_{e,x}$ from an X-ray spectrum.  That is accomplished by folding a theoretical emission spectrum through the response function of an X-ray telescope and minimizing chi-squared between the model and the observation.  To first order, T$_e$ comes from the continuum shape, $\rm e^(-h\nu / kT)$, where h$\nu$ is the photon energy and kT is the thermal energy.  Several factors can complicate the T$_{e,x}$ measurement.  It is necessary to separate the shocked ISM or CSM gas from shocked SN ejecta, which are heated by a separate reverse shock.  The separation requires good spatial resolution, and that is particularly challenging for SNRs in the Magellanic Clouds.  We use temperatures from analyses that were specifically meant to pertain to shocked ISM gas.   The X-ray emitting plasma is not in ionization equilibrium.  Models of shock-heated gas are available, but the ionization timescale is an additional free parameter in the fit.  Elemental abundances introduce still more free parameters, because sputtering gradually liberates refractory elements from dust grains in the hot postshock gas.  In addition, there are many cases where a single temperature model does not provide an adequate fit, so the T$_{e,x}$ determination is ambiguous.  Finally, in some cases X-ray synchrotron emission contaminates the spectrum of the thermal emission.  That being said, observed temperatures of order 1 keV or less are easily distinguished from the higher temperatures that the faster shocks would produce if they were in thermal equilibrium. 

The third diagnostic is measurement of the shock speed based on the proper motion of shocks in an SNR whose distance is known, V$_{pm}$.  Proper motions are mostly measured with \Ha\/ filaments because they are sharp, and because high spatial resolution can be achieved, for instance by the Hubble Space Telescope (HST).  SNR filaments are tangencies of a line of sight to the thin sheet of shock-heated gas \citep{hester87}, so their motion is perpendicular to the line of sight, and the proper motion should correspond to the actual shock speed.  However, \citet{shimoda15} simulated a SNR shock in an inhomogeneous medium, and they found that the rippled shock front could put some energy into turbulence rather than heat.  In their models, comparison of V$_{pm}$ with T$_p$ could overestimate the energy in cosmic rays or in electron thermal energy by up to 40\%.    Proper motions can also be measured from the expansion of a remnant in Chandra X-ray images, provided that a long temporal baseline is available and that there are enough point sources in the field for image registration, or from expansion seen in IR images.  In many cases, uncertainty in the distance to the SNR is the main limitation.  We will consider a few Galactic SNRs whose distances are well-established, along with SNRs in the Magellanic Clouds.    

\begin{table*}
\centerline{Table 2}
\centerline{Equilibration from \Ha\/ profile and T$_{e,x}$ from X-rays }
\vspace*{0.1in}
\begin{centering}
\begin{tabular}{l c c c c c c c c l }
\hline\hline
SNR & Pos & N$_{avg}$ & V$_s$ & T$_p$ & T$_e$ & T$_{Coul}$ & \teti & Ref   \\
     &    &    & \kms &  keV  & keV & keV & & \\
\hline
\\
SN1006 & NW & 1 &  2800 & 18.3 & 0.90 & 0.93 & 0.05 & W13, N13 \\
Tycho  &  &     \\
Kepler & K2-K4 & 3 & 1250-1850 & 5.2-7.2 & 0.68-1.92 & 1.24-1.60 & 0.19 & B91, K15     \\
RCW 86 & SEouter & 1 & 1530 & 2.3 & $<$0.7  & $<$0.3 & 0.1-0.4 &  H11, M14  \\
Cygnus Loop &   North  & 1 & 300-325 & 0.21 & 0.17 & 0.10  & 0.8    & M14, S09     \\
0509-67.5     &    \\
0519-69.0     &   & 4 & 1300-2800 & 2.9-18.3 & 1.43 & $<$2.2 & 0.13 & S16, H18 \\
0548-70.4   & - & 1 & 720-800 & 1.2 & 0.42  & 0.54 & $<$0.42 & S16, S91     \\
DEM L71  & - & 7 & 450-950 & 0.39-1.72 & 0.26-0.43 & 0.5 & $<$0.23 & R03, A22   \\
1E102.2-7219    &   \\
 \\
\hline
\vspace*{0.1in}

\end{tabular}

The key to the references is given with Table 1.
\end{centering}

\end{table*}

\section{Four \teti\/ Determinations}

Different methods can be used to to determine \teti\/ depending on the data available.  Here we describe four methods, their uncertainties and the trends of \teti\/ with shock speed.

\subsection{\Ha\/ Line Profiles: Line Width and \ibin }

The combination of the broad component line width and the broad-to-narrow component intensity ratio of a Balmer line filament, sometimes called a nonradiative shock, can be used to determine \teti .  This method has the advantage that a single measurement of the H$\alpha$ profile provides both the shock speed and \teti .  It is also attractive because to first order,  the profile is determined by just three well-understood atomic rates; the ionization, charge transfer and excitation rates of H I.  An unusual aspect of the atomic rates is that excitation and ionization by protons are important for shock speeds above about 1000 \kms.  A second unusual feature is that the narrow component neutrals move at 3/4 V$_S$ relative to the postshock protons, so that a 2D integration over the relative velocity distribution weighted by the velocity-dependent cross section is required \citep{chevalier80, ghavamian01, heng07, vanadelsberg08, morlino13}.  The asymmetry in the velocity distribution leads to polarization of \Ha\/ \citep{laming90, sparks15, shimoda18}.  The determination of the broad component width and \ibin\/ from the line profile is generally robust unless the spectral resolution is inadequate or the broad component is not Gaussian \citep{raymond10}.

There are two potential difficulties involving the narrow line intensity, however.  First, most of the excitations to the 3p level produce Ly$\beta$ photons.  The branching ratio for H$\alpha$ is 0.12, so a Ly$\beta$ photon will typically convert to H$\alpha$ after about 8 scatterings.  In principle the optical depth to the narrow component Ly$\beta$ photons in the upstream direction is effectively infinite, while in the downstream direction it is typically a few \citep{chevalier80, ghavamian99}.  This leads to a situation between Case A, where all the Ly$\beta$ photons escape, and Case B, where all the Ly$\beta$ photons are converted to H$\alpha$.  The radiative transfer has been computed for various assumed line widths and preshock neutral fractions, and the conversion efficiency seems not to be too sensitive to those parameters \citep{chevalier80, hester94, ghavamian01}.  However, there can be a broader component to the narrow line \citep{morlino13, knezevic17}, and the precursor may accelerate the gas away from the preshock rest velocity, which might reduce the efficiency of conversion of Ly$\beta$ to H$\alpha$.  It should be noted that \ibin\/ can also depend on the preshock neutral fraction, f$_0$.  \citet{morlino13} show results for a reference value f$_0$ = 0.5.  Since f$_0$ will not generally be much larger than that because of photoionization by He I and He II photons from the shock, and it will not usually be much smaller because that would make the Balmer emission very faint, we use that value.  The two exceptions are SN1006, where f$_0$ is about 0.1 \citep{ghavamian02} and knot "g" of Tycho, where f$_0$ may be as large as 0.8 \citep{ghavamian00}.   

The second potential difficulty is more complex.  Gas upstream of a shock will be heated and ionized by ionizing photons from the SNR, and it may produce significant H$\alpha$ emission.  Photoionization precursors are likely to be arcminutes thick for Galactic SNRs \citep{ghavamian00, medina14}.  In principle, emission from such a precursor is subtracted along with the Geocoronal and Galactic \Ha\/ backgrounds, but it may vary on smaller scales and be difficult to subtract.  

There are also precursors associated with cosmic ray acceleration and with broad component neutrals that overtake the shock and deposit their energy upstream \citep{hester94, smith94, raymond11, morlino12, ohira12}.  In Galactic SNRs these cosmic ray or neutral return precursors are on the order of arcseconds  thick \citep{lee10, katsuda16}.  Therefore, depending on the size of the region observed and on the seeing, a larger or smaller contribution from the precursor could be present.  So far, there are few calculations of the H$\alpha$ emission from the precursor, but \citet{morlino12} compute the emission from the precursor and \ibin\/ for shocks faster than 1000 \kms .  The precursor not only contributes to the narrow line emission, but ionization within the precursor can reduce the number of neutrals that reach the shock and undergo charge transfer to produce the broad component.  

The \citet{morlino12} calculations assume Case B, so they somewhat overestimate I$_N$, and the results depend on the uncertain degree of electron-ion equilibration in the precursor.  In addition, \citet{morlino12} did not include helium.  For very fast shocks, this is appropriate, since there is little temperature equilibration among different ion species, but at lower shock speeds the helium ions can significantly heat the protons \citep{raymond15, raymond17}. With those caveats, we use Figure 16 of \citet{morlino12} to estimate \teti\/ from \ibin\/ for shocks faster than about 1000 \kms\/ and \citet{ghavamian01} for slower shocks.  The contribution of the precursor to the \Ha\/ narrow component can be assessed to some extent from intensity of associated [N II] emission \citep{medina14}.

The resulting estimates of \teti\/ are shown in Table~1.  The columns give the name of the SNR, the position within the SNR if one is given and if the measurement is not averaged over too large a region, the number of positions averaged (N$_{avg}$), the broad and narrow component line widths, the broad to narrow intensity ratio, and the shock speed and \teti\/ derived from those values. The key to the references in this and other tables is included.  There is a definite trend of high values of \teti\/  in shocks slower than about 1000 \kms , while faster shocks show values below about 0.2.  This is similar to the results compiled by \citet{ghavamian07b}.

\begin{table*}
\centerline{Table 3}
\centerline{Equilibration from V$_{pm}$ and T$_{e,x}$ }
\vspace*{0.1in}
\begin{centering}
\begin{tabular}{l c c c c c c c c l }
\hline\hline
SNR & Pos & N$_{avg}$ & PM & D  & V$_{SHOCK}$ & T$_e$ & T$_{Coul}$ & \teti &Ref   \\
    &    &  & \arcsec/yr & kpc & \kms & keV & keV & & \\
\hline
\\
SN1006 & SE & 8 & 0.50-0.70 & 1.85 & 4500-6300 & 1.60-2.10 & 0.57 & 0.02 & Mi12 \\
Tycho  & west & 6 & 0.28 & 3 & 4000 & 1.7-2.4 & 0.51 & 0.027  & H02, W16  \\
Kepler &          &     \\
RCW 86 & SEouter & 1 & 0.13 & 2.2-2.6 & 1530 & $<$0.7  & $<$0.3 & 0.1-0.4 &  H11, M14  \\
Cygnus Loop &   North  & 14 & 0.069-0.138 & 0.725 & 300-470 & 0.17 &  0.08-0.16 & 1.1 & S09     \\
0509-67.5     &    \\
0519-69.0     &  3 & 1 & 0.011 & 50 & 1470-3700 & 1.4 & 1.9 & 0.04-0.07 & S16, H18, W22 \\
0548-70.4     &      \\
DEM L71  &    \\
1E102.2-7219    & - & 5 & 0.0055 & 60 & 1610$\pm$370 &  0.66 & 1.72  & 0.07-0.24 &   X19  \\

\\
\hline

\vspace*{0.1in}

\end{tabular}

The key to the references is given with Table 1.
\end{centering}
\end{table*}

\subsection{\Ha\/ Line Width and T$_{e,x}$ from X-ray Spectra}

This method was used by \citet{rakowski03} in a study of DEM L71, where they found that
the faster shocks (700-1000 \kms ) were consistent with no electron-ion equilibration, while the slower ones (400-600 \kms ) were consistent with full equilibration at the shock.  This method does not suffer from problems with background subtraction and contributions to the narrow component from a shock precursor.  Nor does it depend on the distance to the SNR as the methods that use proper motion velocities do.  The method is almost model-independent, it that it uses a direct measure of T$_e$ from X-rays and an almost direct measure of T$_p$ from the H$\alpha$ line width.  It is somewhat model dependent in that the \Ha\/ line width is given by an integral over the charge transfer cross section times the relative velocity distribution \citep{chevalier80}, which is further complicated if a particle goes through multiple charge transfer interactions \citep{heng07, vanadelsberg08, morlino13}.  Here, for shocks faster than about 1000 \kms \/ we use the models of \citet{morlino13} to infer a shock speed from the line width, then use that shock speed to compute T$_p$ from the Rankine-Hugoniot jump conditions. 

The main limitation of this method is that the X-ray spectrum pertains to a relatively thick region behind the shock as compared to the region where the \Ha\/ line is formed.  That means that there is time for Coulomb collisions to transfer energy from ions to electrons, so that the temperature from X-rays is an upper limit to electron heating in the shock itself.  If the X-ray temperatures are determined from non-equilibrium ionization (NEI) fits, the product of density times time, $n_et$, is also determined.  We use that with the proton temperature and the assumption that the electrons are initially cold to compute T$_{Coul}$, the temperature that would be produced by Coulomb collisions alone.  In most cases, this is less than the value of T$_e$ determined from X-rays, but not insignificant. In a few cases it is larger than the measured X-ray temperature, indicating a problem with either the temperature or the value of $n_et$ from the fit.

The resulting estimates are shown in Table 2.  The electron and proton temperatures from the X-ray spectra and \Ha\/ profiles, are given, along with the implied shock speed and the value of T$_{Coul}$ which could be reached by Coulomb collisions alone, and finally \teti .  As with the determinations from the \Ha\/ profile, there is a clear trend of decreasing \teti\/ with shock speed from near 1 in the slow shocks in the Cygnus Loop to 0.05 in the fast shocks in SN1006.

\subsection{V$_{pm}$ and T$_{e,x}$ from X-ray spectra}


The shock speed and the Rankine-Hugoniot jump conditions predict the average temperature, so a measurement of T$_e$ implies a value of T$_p$, and therefore \teti .  We use the jump conditions for positions far upstream and far downstream of the shock.  The effects of ionization of neutrals are discussed in Secion 6.1.  The main complication is that shocks in an inhomogenous gas become rippled and partly oblique.  This can result in a somewhat lower postshock temperature than would be expected from a shock speed derived from the proper motion \citep{shimoda15}.    

The second complication is that if particle acceleration is efficient, it reduces the thermal energy available.  The reduced proton thermal energy can lead to an overestimate of \teti .  The shocks we discuss here do not accelerate particles very efficiently based on a lack of synchrotron X-ray emission at the positions we consider, though other sections in some of the SNRs do have synchrotron X-ray filaments.  Many do produce nonthermal radio and gamma-ray emission, in some cases by reacceleration of existing cosmic rays \citep{tutone21}.  Here, we assume that cosmic rays absorb a negligible fraction of the shock energy, and we return to that assumption in the discussion section, where we summarize estimates of the fractions of shock energy in cosmic rays.   


We consider the rather slow shocks in the Cygnus Loop (300-400 \kms; \citet{salvesen09}),
faster shocks in SN1006 ($\sim 5000$ \kms; \citet{miceli12}), and shocks in Magellanic Cloud supernova remnants, where proper motions show speeds of 1400-3700 \kms\/ \citep{hovey18, xi19, williams18} and electron temperatures have been measured in the outermost regions of the forward shocks chosen to exclude emission from the ejecta \citep{schenck16}. 

We note that, as for the T$_{e,x}$ vs. T$_p$ from the \Ha\/ line width method, the measurement of T$_e$ pertains to a region that is around $10^{17}~\rm cm$ or more thick, so that Coulomb collisions can transfer a significant amount of heat from protons to electrons.  Therefore, we again include T$_{Coul}$ in Table 3, an estimate of T$_e$ that would be reached by Coulomb collisions alone if the electrons were initially cold.

Table 3 shows the proper motion shock speeds, the electron temperatures derived from X-rays, the temperatures that would be reached by Coulomb collisions alone, and \teti .  Again, the trend of \teti\/ decreasing with Mach number is similar to that obtained from \ibin .

\subsection{V$_{pm}$ vs T$_p$ from \Ha\/ broad component}

Some reliable shock speeds have been obtained from measured proper motions and SNR distances, along with \Ha\/ profiles.  The Rankine-Hugoniot jump conditions predict the total thermal energy, and since the proton temperature can be inferred from the \Ha\/ broad component width, we can determine \teti .  As discussed above, the \Ha\/ broad component width is not exactly the same as the proton thermal width, but it is determined by integrating over the cross section times the relative velocity of the neutrals and protons \citep{chevalier80, heng07, morlino13}.  Especially at shock speeds above 2000 \kms , the line width is smaller than the proton thermal width.  Therefore, we use V$_{pm}$ the models of \citet{morlino13} to derive T$_p$ from the \Ha\/ width.

\begin{table*}
\centerline{Table 4}
\centerline{Equilibration from V$_{pm}$ and T$_p$ from \Ha\/ width}
%
\vspace*{0.1in}
\begin{centering}
\begin{tabular}{l c c c c c c c l}
\hline\hline
SNR & Pos & N$_{avg}$ & PM & D  & V$_{pm}$ & V$_{FWHM}$ &  \teti  & Ref   \\
    &      &   & \arcsec/yr & kpc & \kms & \kms & \\
\hline
\\
SN1006 & NW & 11 & 0.37-0.45 & 1.85 & 3000 & 2900 & $<$0.1 & N13, W03, R17  \\
Tycho  & knot "g" & 1 & 0.20 & 2.3-3.7 & $>$2200 & 1800 & $<$0.4 & K78, G01      \\
Kepler &    &   &    &  &  &  &  &      \\
RCW 86 & SEouter & 1 & 0.13 & 2.2-2.6 & 1530  & 1120  & 0.1-0.4 &  H11, M14\\
Cygnus Loop & N & 3 & 0.069-0.11 &  0.725 & 300-360 & 203-347 & 1.0    &  S09, M14   \\
0509-67.5     &  NE & 3 & 0.029 & 50 & 6900 & 7800 & $<$0.1 & H18, M13  \\
0519-69.0     & 2-S & 1 & 0.011 & 50 &  2300-3300 & 3200-4000 & $<$ 0.20 & H18, M13  \\
0548-70.4     &    \\
DEM L71  &    \\
1E102.2-7219    &    \\

\\
\hline
\vspace*{0.1in}

\end{tabular}

\end{centering}

The key to the references is given with Table 1.
\end{table*} 

Table 4 gives the proper motions and inferred shock speeds, along with the FWHM of the broad \Ha\/ component and \teti . Once again, \teti\/ tends to decline with shock speed, from approximatedly 1 for the $\sim$400 \kms\/ shocks in the Cygnus Loop to less than 0.1 in shocks faster than 3000 \kms\/ in SN1006 and 0509-67.5.

\subsection{Evidence from other Observations}

Ultraviolet spectra are only available for a few SNRs, but they support the estimates of \teti\/ given above.  First, \citet{laming96} used the intensity ratios in SN1006 to show that \teti $<$0.05.  Second, kinetic temperatures of H, He, C, N and O have been measured in a handful of SNRs that are observable in the UV and in \Ha\/ \citep{raymond95, korreck04, ghavamian07a, raymond15, raymond17}, as well as X-rays \citep{miceli19}.  They show fairly complete equilibration in the Cygnus Loop, where  the shocks are around 350 \kms, and mass proportional temperatures (no equilibration) in SN1006, where the shocks are faster than 2500 \kms .  It is possible that the wave modes responsible for sharing energy between electrons and ions could be different from those that redistribute energy among the different ion species.  However, the drop in ion-ion thermal equilibration seems to mirror that in electron-ion equilibration.

Another estimate of \teti\/ is given by \citet{matsuda22}, who find \teti\/ $<$ 0.15 in a knot on the NW limb of Tycho based on the X-ray temperature increase over 15 years and a shock speed of $>$ 1500 \kms .  There is also an estimate of \teti\/ for the reverse shock in Tycho by \citet{yamaguchi14}, who used the centroid of the Fe K$\alpha$ line to determine the ionization state of iron.  They found 0.003 $\leq$ \teti\,$\leq$ 0.02 for the reverse shock.  Most recently, \citet{ellien23} were able to separate thermal from non-thermal X-ray emission at five positions in the South, West and Northeast regions of Tycho, and the 2.0-2.6 keV electron temperatures they derive are much less than the 10-15 keV predicted from the shock speeds of \citet{williams16} and the assumption of equal electron and ion  temperatures. That indicates that shocks that produce strong synchrotron X-rays also show \teti $<$ 0.1 of the energy going into cosmic rays and magnetic field amplification is not too large (see Section 6.2). 

We also note that, while Table 1 presents numbers for a 600 \kms/\ shock in RCW86 from \citet{ghavamian01}, which was analyzed in the greatest detail, several other positions in RCW86 were presented in \citet{ghavamian99}.  Excluding 3 positions contaminated by radiative shocks, there were 6 other positions whose line widths and \ibin\/ ratios were similar to those in the one in Table 1, so they would have similar shock speeds and values of \teti\/ around 0.5.  There were also 2 positions showing line widths of 325 \kms and \ibin = 1.0$\pm$0.2, indicating shock speeds of around 350-400 \kms\/ and \teti\/ $>$ 0.6.   These faster shock positions are in good agreement with the RCW86 value shown in Table 1, while the slower shocks agree with the Cygnus Loop value.  Overall, these additional points support the trends in Table 1 and Figure 3.

This paper has concentrated on shocks in SNRs, but pure Balmer line spectra are also observed in the bow shocks driven through the ISM by some pulsars.  \citet{romani22} measured an H$\alpha$ profile in the bow shock of PSR J1959+2048.  The shock speed is probably about 150 \kms , but it is not well determined.  The \ibin\/ ratio is about 4, which would indicate a low value of \teti, probably less than 0.2.  This is much different than the Cygnus Loop shocks, where \ibin\/ is about 1.  One possible cause of this difference is emission from a precursor in the Cygnus Loop.  Another is that the \teti\/ is lower in the bowshock because it is slower. Other possible causes include that the lower temperature in the bow shock favors excitation to the 3s and 3d levels of hydrogen as opposed to 3p and that the smaller width of the broad component permits some conversion of Ly$\beta$ to H$\alpha$ in the broad component.  The latter might make \teti\/ as high as 0.5 consistent with \ibin\/ = 4.  If the shock speed is significantly less than 150 \kms, the lower postshock temperature would reduce the ionization rate, which could strongly increase \ibin.  Since the shock is oblique, the effective shock speed could be smaller than 150 \kms .

\section{Discussion}

All four estimates of \teti\/ in SNRs agree that the electron-ion temperature ratio declines sharply with shock speed.  The discrepancies between SNR observations and solar wind measurements appeared when \teti\/ was plotted against V$_S$, M$_A$ or M$_{ms}$  in Figure~\ref{ghavamian}, which is adapted from Figure 7 of \citet{ghavamian13}.  Measurements at the bowshocks of Earth and Saturn appeared to disagree when plotted against V$_S$, but agreed nicely with each other when plotted against M$_A$ or M$_{ms}$, while the discrepancy with SNR shocks was magnified by plotting against M$_A$ or M$_{ms}$.  

The plots just described assumed generic sound speeds and Alfv\'{e}n speeds for the SNRs of 11 \kms\/ and 9 \kms, respectively.   Here, we consider the possibility that electron heating occurs only in the narrow collisionless shock, while protons are also heated in the precursor.  In that case, the relevant sound speed is the sound speed in the precursor, which is given by the narrow line width, rather than the sound speed in the ambient ISM.  Therefore, we use the widths of the \Ha\/ narrow components to determine temperatures and sound speeds.  Because nearly all of the observed shocks are Balmer line filaments, we assume that the gas is 50\% ionized, except for SN1006 and 1E102.2-7219, which have low neutral fractions.  As shown in Table 5, the narrow component widths vary, and the corresponding temperatures range from T$_4$ = 1.0 to 4.6, where T$_4$ = T/10,000 K.  We note that Tycho's SNR shows both a narrow and an intermediate component that account for 35\% and 65\% of the `narrow' component.  We have taken the weighted average of the corresponding temperatures in quadrature.

The Alfv\'{e}n speed is even more problematic, given that neither the ambient magnetic field nor the compression in the precursor is known.  For Milky Way SNRs we assume a field strength of 10 $\mu$G \citep{sofue19}.  Kepler's SNR and RCW 86 are expanding into their own circumstellar shells, which could have very low magnetic fields or very high fields due to compression, and the preshock densities vary enormously around their peripheries.  For lack of reliable numbers, we do not compute Alfv\'{e}n or magnetosonic speed for those remnants.  For the LMC and SMC we also assume 10 $\mu$G based upon \citet{mao12}, \citet{seta21}, \citet{seta22} and \citet{livingston22}.  These are, of course, average magnetic fields, so the local field will be different.  To obtain Alfv\'{e}n speeds, we use preshock densities generally derived from X-ray observations and an assumed compression ratio of 4 at the shock.  As seen in Table 5, the preshock densities vary by over an order of magnitude.  The Alfv\'{e}n speed may change in the precursor, but as the plasma and the perpendicular component of the field are compressed together, the change in V$_A$ should not be drastic.

%
%





\begin{table*}
\centerline{Table 5}
\centerline{Sound Speed and Alfv\'{e}n Speed }
%
%
\vspace*{0.1in}
\begin{centering}
\begin{tabular}{l c c c c c c c l}
\hline\hline
SNR & Pos  & V  &FWHM$_{narrow}$ & T$_4$ & n$_0$  & C$_S$ & V$_A$ & Ref  \\
     &      & \kms & \kms & 10$^4$ K & \cmcube & \kms & \kms & \\
\hline
\\
SN1006      & Avg      & 3000  & 22 &  1.0 & 0.1 & 16 & 63  &  B19 \\
Tycho       & knot "g" & 2200  & 49,188 & 28. & 1.0 & 65 & 20 & K17, G00  \\
Kepler      & Avg      & 2000  &   42 & 3.5 & - & 23 & - & S03     \\
RCW 86      & SW       & 560   &   36  & 2.6 & - & 20 & - & S03, S17 \\
Cygnus Loop & Avg N    & 320   & 35 & 2.5 & 0.20 & 20 & 44 & M14, S09 \\ 
0509-67.5   & Avg      & 6000  & 28 & 1.6 & 0.4 &  16 & 31 & S91 S21   \\
0519-69.0   & Avg      & 2500  & 40 & 3.2 & 1.5 &  22 & 16 & S94, K10, W22    \\
0548-70.4   & Avg E    & 760   & 48 & 4.6  & 0.4 & 26 & 31 & S91, S94, B06    \\
DEM L71     & Avg      & 800   & 37 & 2.8  & 0.6 & 21 & 26 & S94, B06    \\
1E102.2-7219    & Avg  & 1600  & - & 1.0 & 1.6 & 12 & 16 & X19   \\

\\
\hline
\vspace*{0.1in}

\end{tabular}

\end{centering}

The key to the references is given with Table 1.
\end{table*} 

\begin{figure*}
\begin{centering}
\includegraphics[width=6.65in]{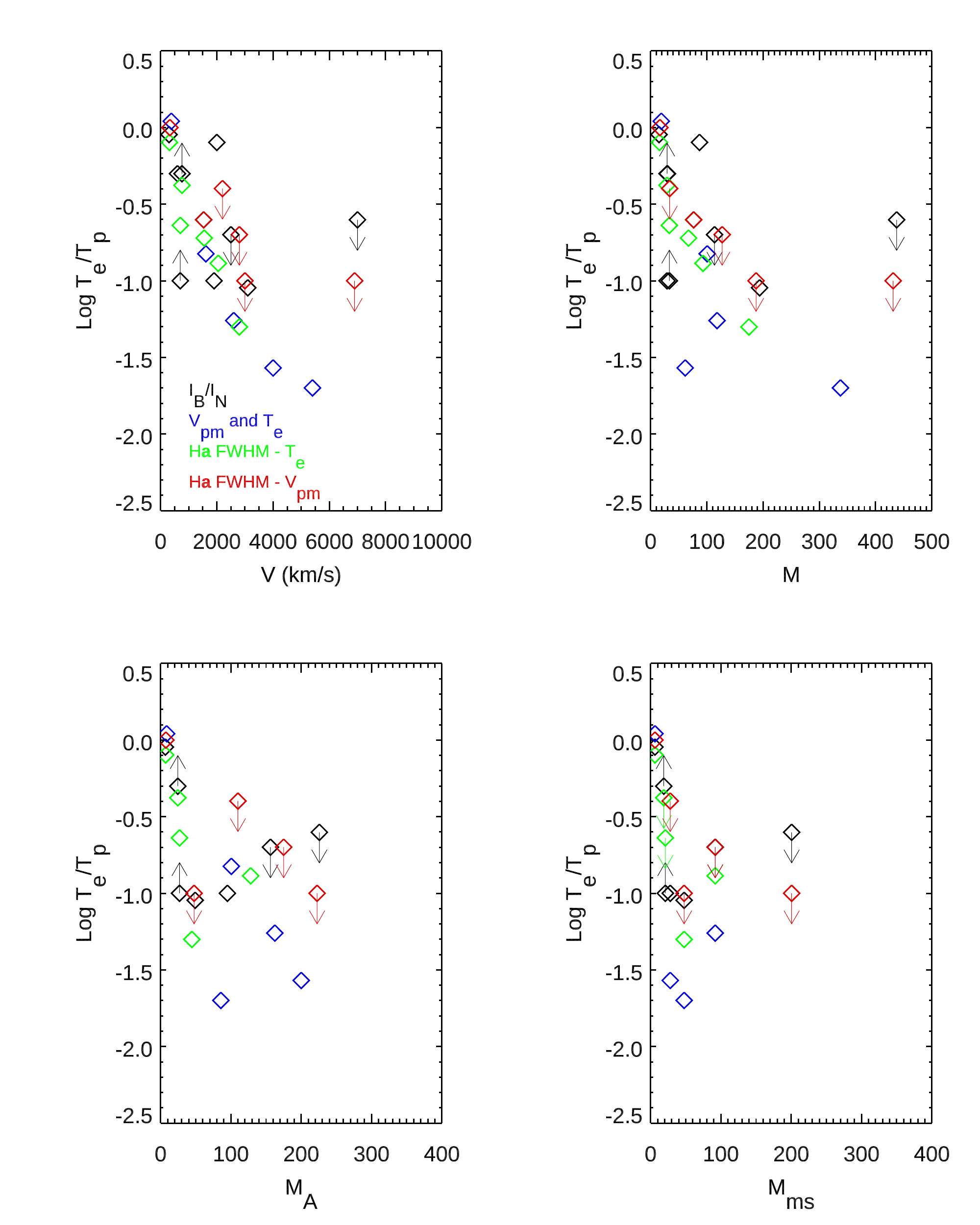}
\caption{Four estimates of Log \teti\/ plotted against shock speed, sonic Mach number, Alfv\'{e}n Mach number and magnetosonic Mach number.  At most one value is given for each SNR for each method of estimations, and it is either an average or the best-determined value, as discussed in the Appendix. The estimation method for \teti\/ is indicated by the color.  Note that the X-axis scales are different.
\label{teti}
}
\end{centering}
\end{figure*}

Based on the estimates of sound speed and Alfv\'{e}n speed in Table 5 and the values of \teti\/ in Tables 1-4, we plot in Figure~\ref{teti} the equilibration as a function of shock speed, sonic Mach number, Alfv\'{e}n Mach number and magnetosonic Mach number, using different colors for the different methods of estimation.  To reduce the clutter, we plot only a single point for each SNR for each method of estimation, generally an average if a range of values is present, but in some cases such as RCW 86, the best determination for that SNR.  We also introduce slight shifts along the X-axis in cases where a symbol in one color would cover that in another color.

Figure~\ref{teti} shows that the trends of \teti\/ with shock speed shown in Figure 7 of \citet{ghavamian13}  (our Figure 1) hold, but that the Mach numbers are considerably smaller.  This is due to the generally higher temperatures assumed here based on the narrow component line widths, to the generally higher Alfv\'{e}n speeds due to smaller preshock density estimates, and to an assumed 10 $\mu$G magnetic field.  However, the changes in Mach number, Alfv\'{e}n Mach number and magnetosonic Mach number are not nearly enough to bring the SNR results into agreement with the solar wind results (\teti $\sim$0.15 at V$\sim$400 \kms , M$_A \sim$10, M$_{MS} \sim$7), or with the PIC prediction that \teti\/ rises to about 0.1 to 0.3 when M$_A$ exceeds about 20 to 30.

There are, or course, still ways in which our Mach numbers could be overestimates.  For very short period Alfv\'{e}n waves, the density of the ionized gas rather than the total density should be used.  The ionization fraction is about 90\% in SN1006 \citep{ghavamian02}, and it cannot be much less than about 0.5 because the shock itself produces ionizing photons \citep{medina14}.  We also use mean magnetic fields to estimate the Alfv\'{e}n speed, and the local fields can be higher or lower.  We also use the pre-shock temperature estimated from the narrow component line width, and it is possible that protons are further heated in the precursor after the last charge transfer (on average) communicates the proton temperature to the neutrals.  It is unlikely that these effects would reduce the Mach numbers by more than a factor of 2.
 
 We therefore consider some ways in which the comparison among SNR, solar wind and PIC simulated shocks might be comparing different things.  Then, we consider possible differences in the physics.

\subsection{Different Length Scales}

One obvious difference between observations of SNR shocks, the measurement of solar wind shocks and the PIC simulations is the scale.  The latter two pertain to a few hundred times the ion skin depth, which is on the order of $10^8$ cm in the interstellar medium.  The \Ha\/ profile in an SNR shock is formed over an ionization length scale, typically $10^{15}$ cm, and X-ray spectra are generally resolution-limited to scales of $10^{17}$ cm.  Thus if a shock has a precursor (either a cosmic ray modified shock or a return neutral precursor), the solar wind and PIC results might pertain to the subshock, while the SNR results pertain to  the entire structure.  Similarly, if postshock cooling or other processes affect the plasma between the shock jump and the \Ha\/ formation region, that could affect the comparison.  The comparison of laboratory experimental data with SNR observations suffers from similar challenges due to the small scales of the shock evolution captured in the experiments.  Nevertheless, laboratory experiments can play an important role in benchmarking the results of numerical simulations, which have sometimes been limited by reduced mass ratios and dimensionality. This will strengthen the understanding of the plasma microphysics and its connection to \teti\/ in collisionless shocks.

{\it Cosmic ray precursor:}  A cosmic ray precursor requires magnetic turbulence to reflect accelerating particles back toward the shock, but it is uncertain how effectively the turbulence dissipates or whether it heats ions or electrons more effectively \citep{morlino12}.  The magnetic turbulence in the precursor will contribute to the narrow component line width, causing an overestimate of the proton temperature and sound speed and an underestimate of the sonic Mach number.  By the same token, the narrow component width provides an upper limit on the level of turbulence in the precursor, which is related to the amplification of magnetic fields discussed below.

{\it Return neutral precursor:}  As was mentioned in connection with the narrow component flux, a neutral return precursor, in which some of the broad component neutrals overtake the shock front and deposit their energy upstream, is also expected \citep{hester94, raymond11, morlino13}.  When a neutral overtakes the shock, it has at least the average energy of the shocked protons, and when it is ionized by charge transfer or by collision with an electron, it becomes a pickup ion similar to those seen in the solar wind, but with an energy of 0.5 to 200 keV for the shock speeds considered here.  The pickup ions tend to be swept back through the shock, and what fraction of their energy is deposited upstream is a complex question.  In a perpendicular shock with modest magnetic turbulence, the pickup ions form a ring beam in velocity space, and that very quickly scatters into a bispherical shell distribution in velocity space \citep{williams94, isenberg96}, transferring some of the kinetic energy to Alfv\'{e}n waves in the process. The amount of energy deposited depends on $\Theta_{BN}$, the plasma $\beta$ and on the neutral fraction \citep{raymond08}.  It is likely that protons are heated more effectively than electrons, but at the low temperatures in the precursor, Coulomb collisions may transfer energy to the electrons.     

{\it Postshock electron heating and cooling:}  Another possible factor is that electrons are cooled behind the shock.  Some energy is required to ionize the atoms, and some energy is lost by radiation over the region where the Balmer lines are produced, mostly in the form of H I Lyman lines and He I and He II emission lines.  The ionization and radiative loss energies involved could cool the electrons by removing thermal energy, but the amount is equivalent to reducing the effective shock speed to $v^{\prime^2} = v^2 - 72^2$ (\kms)$^2$\/ for a shock in fully neutral gas \citep{cox85}, and that is negligible for the shocks we consider here.  In addition, ionization of neutrals also liberates electrons, so the thermal energy that electrons gain at the shock is shared among more particles.  The neutral fraction can be as high as 50\% if there are no sources of ionizing radiation in the neighborhood other than the shock itself, so the liberation of neutrals downstream could cool the electrons to half their value at the shock.  That could be a significant factor for the SNRs showing \teti\/$\sim$ 0.5, such as RCW 86 or 0548-70.4.  The SNR with the most reliably known preshock neutral fraction is SN1006, where only 10\% of the hydrogen is neutral \citep{ghavamian02}, and there the electrons created by ionization downstream would have only a 10\% effect on the temperature.  Another possible cooling process is adiabatic expansion, but that would tend to occur on large scales, over which Coulomb collisions could heat the electrons.

Another possible mechanism for heating the electrons downstream is the cosmic ray postcursor discussed by \citet{diesing21} and \citet{wilhelm20}.  Here, magnetic turbulence in the postshock region is maintained for a thickness comparable to that of the cosmic ray precursor.  If that turbulence is dissipated by reconnection, it could preferentially heat the electrons.  That could help to explain why SNR shocks at Mach numbers around 10-15 show substantial equilibration, but it would aggravate the discrepancy between PIC predictions and SNR observations of T$_e$ at high Mach numbers.  The energy available is discussed below in connection with cosmic rays. 

\subsection{Jump Condition Energetics; Cosmic Rays and Magnetic Fields}

The \teti\/ estimates based on V$_{pm}$ rely on the shock jump conditions and  the differences between the inferred total thermal energy and the thermal energies measured for either protons or electrons.  With the exception of the analyses of RCW 86 by \citet{morlino14} and 0509-67.5 by \citet{hovey18}, most of these estimates assume that the fractions of shock energy going into cosmic rays and amplification of magnetic fields, $\epsilon_{CR}$ and $\epsilon_B$,  are small.  If the energy in cosmic rays and magnetic fields is significant, it will affect each estimate of \teti\/ in a different way.  The energy to accelerate cosmic rays comes from the protons, since they carry almost all the kinetic energy that is dissipated. 

The estimates of \teti\/ based in \ibin\/ and on the X-ray spectrum vs. the \Ha\/ line width rely on direct measures of T$_e$ and T$_p$, so they do not depend on assumptions about the fraction of energy in cosmic rays, but the shock speed would be underestimated by a factor $(1-\epsilon_{CR}-\epsilon_B)^{1/2}$. The estimate obtained by comparing $T_{e,x}$ with the proper motion shock speed assumes that the ions have all the dissipated shock energy that is not in the electrons, so that T$_p$ is overestimated and \teti\/ is underestimated if $\epsilon_{CR}$ is significant.  The estimate based on comparing T$_p$ from the \Ha\/ width with V$_{pm}$ assumes that all the dissipated shock energy that is not in the ions is in the electrons, so it overestimates T$_e$ and \teti\/ if $\epsilon_{CR}$ is important.

If SNRs provide much of the energy in cosmic rays, then the fraction of shock energy in cosmic rays must be $\epsilon _{CR}$ = 0.05-0.10 \citep{strong10} on average, though $\epsilon_{CR}$ may be much larger in fast shocks than in slower ones \citep{blasi05}.  We only have detailed estimates for a few SNRs.  \citet{tutone21} find that $\epsilon_{CR}$ is only about 0.02 in the 400 \kms\/ nonradiative shocks in the Cygnus Loop.  \citet{hovey18} find upper limits $\epsilon_{CR} < 0.07/(1-f_0)$ and $\epsilon_{CR} < 0.11/(1-f_0)$, where $f_0$ is the preshock neutral fraction, in 0509-67.5 and 0519-69.0, respectively, with shock speeds spanning 1700 to 8500 \kms .  \citet{morlino14} studied a 1500 \kms\/ shock in RCW 86, and found 0.05 $<~ \epsilon_{CR}~ < 0.30 $.  With the possible exception of some positions in Tycho's SNR, none of the regions studied here show nonthermal X-ray emission.  Therefore, it is unlikely that $\epsilon_{CR}$ is more than 10\% in the regions we consider. The recent work of \citet{ellien23} determined electron temperatures in Tycho's SNR shocks by separating the thermal and nonthermal components of the X-ray spectra.  Their comparison with temperatures determined from proper motion shock speeds indicates that \teti is 0.09-0.13 for $\epsilon_{CR} = \epsilon_B=0.1$.  Thus shocks that produce X-ray synchrotron filaments seem not be behave much differently from shocks that do not.

Cosmic ray acceleration is accompanied by turbulent amplification of the magnetic field.  That can reduce the energy available, but it also provides a nonthermal component to the line width.  It has generally been assumed that kinetic scale turbulence would damp out over a length much smaller than the thickness of the region where \Ha\/ forms, but a postcursor in which cosmic rays and turbulence exchange energy is also possible \citep{wilhelm20, diesing21}.  The energy in magnetic turbulence could reach equipartition with the cosmic ray energy, $\epsilon_B = \epsilon_{CR}$, so it could be as much as 10\% of the energy dissipated in the shock.  It would reduce T$_p$ like the cosmic rays, but it would increase the \Ha\/ line width.  In the reference case of $\epsilon_B = \epsilon_{CR}$ = 0.1, the energy density in the waves is 1/8 the thermal energy density.  For Alfv\'{e}n waves, the magnetic and kinetic energy contributions are equal, so the wave kinetic energy is 1/16 the thermal energy, and if T$_p >>$ T$_e$ the wave velocity amplitudes are 1/4 the proton thermal velocities.  If T$_p$ = T$_e$, the wave velocity contribution would be $\surd 2$ times higher.  However, the slow Cygnus Loop shocks where T$_p$ = T$_e$ show $\epsilon_{CR}$ = 0.02 rather than 0.1 \citep{tutone21}.  If the wave velocity and thermal velocity add in quadrature, the \Ha\/ line widths are only increased by a few percent.  The compilation by \citet{reynolds21} indicates that $\epsilon_B$ is about 10\% on average for Tycho and RCW 86, but those estimates pertain to the X-ray synchrotron filaments rather than the ones studied here.  The values of $\epsilon_B$ from \citet{reynolds21} are much smaller for SN1006 and Kepler.

\subsection{Other Parameters}

Other shock parameters might influence \teti .  The angle between the shock normal and the magnetic field, $\Theta_{BN}$, could well have an effect according to PIC simulations, as discussed in Sections 2 and 3.   There is no reason to believe that the SNR shocks are preferentially either quasi-parallel or quasi-perpendicular, though our avoidance of shocks with X-ray synchrotron emission might bias our sample toward perpendicular shocks according to some models.  The ratio of gas pressure to magnetic pressure, $\beta$, could perhaps have an effect, but the preshock values of $\beta$ should be of order 0.1-1, and the solar wind shocks should be roughly similar. As also discussed in section 2, the plasma $\beta$ seems to have little effect for the modest Mach number shocks, and its importance would be expected to be even smaller in faster shocks.

{\it Subcritical/Supercritical shocks:}  One aspect of the shocks that we have not discussed is their subcritical/supercritical character.  Shocks up to M$_{ms}\sim$2.76 (depending on shock parameters) can dissipate energy efficiently enough to make a smooth, steady shock structure possible \citep{treumann09}.  Faster shocks must be unsteady, and reflection of protons at the shock can generate modified two-stream instabilities (MTSI) that produce plasma waves that could couple electrons and ions \citep{wu84}.  While a supercritical shock may be required to produce some of the wave modes that can transfer energy from electrons to ions,  even laminar shocks can produce whistler wave precursors that could transfer energy \citep{gary85}.  The supernova remnant shocks discussed here are well above the range of subcritical shocks, with the possible exception of the Cygnus Loop shocks, which could perhaps be slower than the ``second whistler critical Mach number"  or the ``third or nonlinear whistler critical Mach number", allowing phase-standing whistlers to create a steady structure \citep{kang19}.  In any case, very weak shocks should converge on \teti$\sim$1 at M$\sim$1 due to adiabatic compression even without electron-ion coupling \citep{vink15}.  However, the Cygnus Loop electrons are much too hot for adiabatic heating to account for T$_e$. 

{\it Effects of neutral atoms:}  The most likely parameter to influence electron heating could be the neutral fraction, which is effectively zero in the solar wind and in the PIC simulations, but is 0.1 or more in the Balmer filaments.  Neutrals could damp plasma waves and perhaps weaken those responsible for transferring energy from ions to electrons.  In general, that is expected to happen for relatively long period waves that resonate with highly energetic particles, so that a significant neutral fraction limits the maximum cosmic ray energy \citep{reville07}, while the ion-neutral damping length is much larger than the wavelengths of waves that would interact with thermal particles.  If the presence of neutrals is important, it would affect all the \teti\/ determinations that involve the \Ha\/ profile.  It would not affect the values of \teti\/ determined by comparing proper motion shock speeds with X-ray determined temperatures in the western side of Tycho or in E0102.2-7219, where no Balmer filaments are seen.

\section{Summary}

We have used used four measured parameters of SNR shocks to determine \teti; the \Ha\/ broad component widths, the \Ha\/ broad-to-narrow intensity ratios, the electron temperatures derived from X-ray spectra and the shock speeds derived from proper motions.  They provide four distinct estimates of \teti\/ whose systematic uncertainties are different, and we have discussed the uncertainties for each method.  We have used published values for the measured parameters, though in some cases we reinterpreted the data using newer models \citep{morlino12, morlino13} or newer distances to individual SNRs.

The result is that the trend of \teti\/ obtained from the \Ha\/ profiles by \citet{ghavamian13} is robust, and the disagreements with measurements of solar wind shocks remain, along with disagreement with PIC simulations.  First, the in situ measurements and PIC simulations indicate that \teti\/ is close to 1 at very low M numbers, but drops to $\sim$ 0.2 around M$_A$ near 15, while the SNR shocks are still at \teti\/ near 1 at M = 15-25.  Second, both PIC simulations and in situ measurements indicate that as the shock speed increases for M above about 20, \teti\/ increases to about 0.1 to 0.3, while in the SNR shocks, \teti\/ continues to decrease to 0.05 or less.

We consider several possible avenues to resolve this discrepancy: the scale lengths of the regions where the diagnostic emission is formed, the presence of energetic particles and magnetic turbulence, the roles of shock precursors and postcursors, the Alfv\'{e}n speeds used to derive Alfv\'{e}n Mach numbers, and the presence of neutrals.  While some of these factors decrease the discrepancies, they do not provide definitive explanations.

Thus, we have utterly failed to resolve the discrepancies among SNR shocks, solar wind shocks, and PIC simulations in electron-ion thermal equilibration.  It is not clear whether the answer lies in more theory or more observations.  Systematic observational studies, in which H$\alpha$ profiles, proper motion measurements and high resolution UV and X-ray spectra of individual shocks are obtained, are needed.  On the theoretical side, simulations that include neutrals and the important plasma processes are needed, both upstream and downstream of the shock jump.



\section{Acknowledgements}
This research was supported by the International Space Science
Institute (ISSI) in Bern, through ISSI International Team project \#520. The effort was supported by HST Guest Observer grant GO-15285.0001\_A to the Smithsonian Astrophysical Observatory.
The work of J.~N. has been supported by Narodowe Centrum Nauki through research project No. 2019/33/B/ST9/02569.  The work of DR was supported by the National Research Foundation of Korea through 2020R1A2C2102800. AT was supported by NASA grant FINESST 80NSSC21K1383.  AB was supported by the German Research Foundation (DFG) as part of the Excellence Strategy of the federal and state governments - EXC 2094 - 390783311.

\appendix

\section{SN 1006}  

SN1006 has been extensively studied.  \Ha\/ is present around the entire circumference, but it is too faint for good line profiles to be obtained except in the NW, where \citet{nikolic13} obtained excellent IFU data. Proper motions have been measured both in \Ha\/ and X-rays \citep{winkler03, miceli12, katsuda13, raymond17}.  The distance to SN1006 has been estimated in a number of ways to be 1.6-2.2 kpc.  High velocity absorption from the unshocked ejecta of SN 1006 has been seen in the spectrum of the Schweizer-Middleditch star, whose Gaia parallax indicates 1.43-2.0 kpc.  A distance less than 1.6 kpc would imply shock speeds too small to account for the observed \Ha\/ line widths. Therefore, we adopt D = 1.85$\pm$0.15 kpc to convert proper motions to shock speeds. X-ray temperatures are taken from \citet{winkler13} in the NW and \citet{miceli12} in the SE.  We note that the regions observed by \citet{miceli12} do not show the relatively bright \Ha\/ seen in the NW, but very faint \Ha\/ is present.  That might indicate a neutral fraction even lower than the value of 0.1 in the NW found by \citet{ghavamian02}.  \citet{miceli12} found values of n$_e$t of only 2-7$\times 10^8~\rm cm^{-3}~s$, so T$_{Coul}$ is only about 1/3 the observed X-ray temperatures.  In the NW, \citet{winkler13} find a higher value of n$_e$t, so Coulomb collisions could provide a substantial fraction of the temperature obtained from the X-ray spectra.  We also note that other measurements of T$_{e,x}$ and V$_{pm}$ yield similar results for \teti\/ \citep{long03, winkler03}.




\section{Tycho's SNR}  

Tycho's SNR has also been studied extensively.  For the \teti\/ estimate from \ibin , we use \Ha\/ widths and broad-to-narrow intensity ratios of \citet{ghavamian01} and \citet{kirshner87}, and we interpret them with the models of \citet{morlino12} and \citet{morlino13}. X-ray temperatures of the shocked ISM are available in several regions from \citet{hwang02}, and while \Ha\/ profiles are not available at those positions, we can compare them with proper motions from \citet{williams16}.  Distance estimates range from 2.3 to 3.7 kpc \citep{chevalier80, black84, albinson86}.  A distance of 2.3 kpc gives a lower limit to the shock speed of 3200 \kms, which leads to an upper limit to \teti\/ of 0.15.  The strongest limit comes from the combination of the X-ray temperatures from \citet{hwang02} with the proper motions on the western rim from \citet{williams16} and a distance of 3 kpc, which gives \teti=0.027.  We note that this region does not show Balmer line filaments, so preshock neutral fraction is small.  The proper motion of knot ``g" of \citet{kamper78} is 0.20$\pm$0.01~\arcsec\/ per year.  With the \Ha\/ width of \citet{ghavamian01} and the distance of 2.3 to 3.7 kpc, it implies \teti\/ $<$ 0.4.  Another spectrum in the knot ``g" area is given by \citet{raymond10}.  Within the uncertainties, the width of the broad line agrees with the other values, but a Gaussian broad component did not provide an adequate fit to the data.  Several interpretations, including multiple shocks within the aperture, a non-Maxwellian velocity distribution, a shock precursor or pickup ions, are possible. 

Another estimate of \teti\/ is given by \citet{matsuda22}. They see a substantial increase in T$_e$ from 0.30 to 0.69 keV in a knot on the NW limb of Tycho between 2000 and 2015, which they attribute to Coulomb heating.  They find \teti\/ $<$0.15 at the shock.

\section{Kepler}  

Individual values of \ibin\/ from \citet{fesen89} and \citet{blair91} vary considerably, and some are inconsistent with the models of \citet{morlino12}.  This is likely to be the result of contamination by radiative shocks.  We exclude position P2D2 of \citet{blair91} because its very large FWHM would imply a shock speed above 5000 \kms , and we exclude positions where \ibin\/ cannot be matched by the models.   The average FWHM and \ibin\/ give \teti\/ = 0.8 if $T_e$ = T$_p$ upstream.  Overall, the profiles are suspect, given the possible contamination by radiative shocks.  For the comparison of T$_{e,x}$ with \Ha\/ broad component line width we use \citet{blair91} and \citet{katsuda15}. \citet{coffin22} have measured proper motions, but the distance uncertainty is large enough that we do not use proper motion velocities.

\section{RCW 86}  

\citet{helder11} studied the electron-ion equilibration in RCW 86 based on the broad component widths and electron temperatures from XMM spectra.  In their sample, the three positions with low uncertainties showed \teti\/ $<$ 0.1, two positions showed \teti\/ $\simeq$0.45, and two positions showed \teti$>$0.58.  One position showed an extremely narrow \Ha\/ width that corresponded to \teti$> 10$, but that is probably a case of a slow shock dominating the \Ha\/ profile while a faster shock produces the X-rays.  The shock speeds ranged from 580 \kms\/ to 1100 \kms , but there was no trend of \teti\/ with shock speed.  Table 1 give the value of \teti\/ from \citet{ghavamian00} based on the \Ha\/ profile.  As described under the corroborating observations, \citet{ghavamian99} presented several other \Ha\/ profiles, and they indicate \teti\/ in accord with values in Table 1.

The distance to RCW 86 is 2.4$\pm$0.2 kpc based on the kinematic distance to the associated atomic and molecular shell \citep{sano17}, so proper motion measurements by \citet{helder13} and \citet{yamaguchi16} can be translated into shock speeds of the thermal filaments of 700-2200 \kms , with speeds of 1200$\pm$200 \kms\/ in the NE and SE that correspond very well with the \Ha\/ broad component widths for \teti =0.4. 

\citet{morlino14} combined the proper motion, \Ha\/ line width and X-ray temperature to obtain tight limits on cosmic ray acceleration efficiency and \teti\/ at one position.  We use their value for Tables 2, 3 and 4.  

\section{Cygnus Loop}   

For estimates of \teti\/ based on the \Ha\/ FWHM and \ibin, we use the observations of \citet{medina14} with the \citet{ghavamian01} predictions for \ibin\/ as a function of \teti .  We exclude a point with a large uncertainty (FWHM=347$\pm$89 \kms) and positions with line widths of 141 \kms   and 178 \kms , which would correspond to shocks slow enough to be radiative.  We average the remaining nine positions and adopt \teti\/= 0.8-1.0.

\citet{salvesen09} also measured electron temperatures from ROSAT spectra.  Two-temperature fits were required, but we use the lower temperatures because that component dominated the total emission.  They are in line with other X-ray temperatures in the northern Cygnus Loop \citep{raymond03, katsuda08, sankrit10}. To estimate the contribution of Coulomb collisions to T$_e$ for the X-ray observations, we assume a preshock density of 0.2 and therefore a postshock density of 0.8 cm$^{-3}$, along with a thickness of $5\times 10^{17}$ cm corresponding to the spatial resolution of the ROSAT spectra.  The \Ha\/ filaments are formed much closer to the shock, around $10^{15}$ cm, so Coulomb collisions can only heat the electrons to about $3\times 10^5$ K in the \Ha -emitting region.  We exclude four outlying points whose apparent speeds were less than 300 \kms, as they would be ineffective in producing X-rays and may be transitioning to become radiative shocks.   The observed X-rays probably originate in faster nearby shocks.  That leaves 14 positions whose speeds range from 300 to 470 \kms , and whose values of \teti\/ range from 0.73 to 1.62 and average to 1.10.   

To determine \teti\/ from the \Ha\/ width and T$_{e,x}$ we combine the \citet{salvesen09} and \citet{medina14} results.  Of the six positions where both are available, we exclude the five with line widths below 250 \kms , leaving only Salvesen's position 8.  It yields a value of \teti\/ of 0.8, but the uncertainty in the \Ha\/ line width is 20\%, and the uncertainty in \teti\/ is correspondingly large.

We also combine the data from \citet{salvesen09} and \citet{medina14} to determine \teti\/ from V$_{pm}$ and the \Ha\/ width. Here, we average three positions, excluding the slowest shocks and the position whose line width is most uncertain and obtain an average \teti\/ of 1.0.

Though the relatively bright filament in the NE first discussed by \citet{fesen82} has been studied extensively \citep{long92, hester94, sankrit00, blair05, katsuda16} we do not include it here because it is partially a radiative shock.

\section{0509-67.5}     

There are several sources for \Ha\/ profiles and proper motions, though we have been unable to find an X-ray temperature derived from fits to a region that definitely corresponds to the shocked ISM \citep{smith94, helder10, hovey18, knezevic21}.   The broad component line width varies around the remnant, ranging from around 3000 \kms\/ in the E and to over 4000 \kms\/ in the NE and NW.  The value of \ibin\/ is 0.06$\pm$0.01 according to the MUSE observations of \citet{knezevic21}.  Though some positions show large uncertainties and provide less stringent upper limits to \teti\/, some positions indicate \teti$<$0.1 and probably lower; see Figure 3 of \citet{knezevic21}.

This SNR also has well-determined proper motions \citep{hovey18}, and the distance to the LMC is known, so reliable shock speeds can be derived.  Very recent proper motion measurements were made from  HST \Ha\/ images \citep{arunachalam22}  and from Chandra images \citep{guest22b}.  They show average shock speeds of 6315 and 6120 \kms , respectively, with speeds as high as 8000 \kms\/ in some regions.  The limit on \teti\/ given by \citet{hovey18} combines the \Ha\/ profile information with the proper motions,
and we use that value for both determinations.

\section{0519-69.0}    

For the value of \teti\/ from \ibin\/ and the \Ha\/ line width we use the value of \citet{hovey18}, which folds in proper motion measurements as well.  We quote 1-$\sigma$ upper limits rather than the 95\% confidence values given by \citet{hovey18}.  We also use the 95\% upper limit from \citet{hovey18} (converted to 1-$\sigma$) for the constraint from the \Ha\/ line width and proper motion velocity.

To determine \teti\/ from X-ray observations and proper motion velocities, we combine the X-ray temperatures of \citet{schenck16} with proper motions from \citet{hovey18} and \citet{williams22}.   The \citet{hovey18} positions correspond better to the \citet{schenck16} extraction regions, while  the \citet{williams22} proper motions should be more accurate because of their longer baseline.  The average electron temperature of 1.4 keV and n$_e$t = $5.0 \times 10^{10}~\rm cm^{-3}~s$ are taken from \citet{schenck16}.  The extraction regions for the X-ray spectrum correspond approximately to regions East, 1-S, 2-S and 1-N of \citet{hovey18}, so we use the average \Ha\/ broad component line width of those regions for comparison.  We use the average proper motion speeds of \citet{hovey18} to predict an average proton temperature of 11.6 keV if there is no equilibration.  More recently, \citet{williams22} measured proper motions in 0519-69.0, and  their regions 5, 6, 16 and 20 correspond approximately to the regions where \citet{schenck16} extracted the X-ray spectrum.  Their velocities give an average proton temperature of 16.7 keV.  We use both to estimate a range \teti\/ = 0.04-0.07.  This is a case where Coulomb heating could account for all the electron thermal energy.  \citet{guest23} have very recently remeasured the X-ray expansion rate, and they find an average value of 4760 \kms, at the upper end of the range reported by \citet{williams22}.


\section{0548-70.4}   

The estimate of \teti\/ from \ibin\/ and the \Ha\/ line width uses the measurements of \citet{smith91}.
Taking an X-ray temperature of 0.42 keV from \citet{schenck16} and a shock speed of 760$\pm$40 from the line width given by \citet{smith91} gives \teti = 0.28-0.67.  However, \citet{schenck16} also give n$_e$t = $2.6 \times 10^{11}~\rm cm^{-3}~s$, which is long enough for Coulomb collisions to heat the electrons to 0.5 keV.  We consider the value of n$_e$t to be the most uncertain of the measurements, and indeed somewhat large for the likely age of the SNR.  Simply taking T$_p$ = 1.02 keV from the \Ha\/ line width and the X-ray temperature T$_e$ = 0.42 keV gives \teti = 0.42, and since Coulomb equilibration is probably significant, we take that as an upper limit.  Proper motions are not available for this SNR.

\section{DEM L71}    

We use the \Ha\/ profile measurements of \citet{smith91} and \citet{rakowski09} to find the values of \teti\/ from \ibin.  \citet{alan22} fit X-ray spectra of the shocked ISM in DEM L71, and we have interpolated among positions where \citet{rakowski09} measured the \Ha\/ line widths.  For seven positions, the broad component widths range from 450 \kms\/ to 950 \kms , with an average of 785 \kms.  The corresponding proton temperatures range from 0.39 to 1.72 keV, while the electron temperatures from the X-ray fits range from 0.26 to 0.43 keV.  The average value for \teti\/ is 0.23.  The values of n$_e$t obtained by \citet{alan22} are larger than $10^{11}~\rm cm^{-3}~s$, so substantial Coulomb heating of the electrons has probably occurred.  We note that \citet{rakowski03} also compared \Ha\/ line widths with T$_{e,x}$ at 5 positions around DEM L71.  Their values of n$_e$t at two positions preclude useful determination of \teti , but 3 other positions with shock speeds from 750 to 1000 \kms\/ show \teti\/ averaging 0.7, where Coulomb collisions might account for perhaps half of T$_e$.  No proper motions are available for this SNR.

\section{N103B}   

\Ha\/ profiles at 4 positions are available from a MUSE data cube, and they indicate shock speeds between 1160 and 3560 \kms, depending how much electron-ion equilibration occurs \citep{ghavamian17}.  Unfortunately, the values of \ibin\/ range between 0.23 and 0.48, and they are all below the range predicted by \citet{morlino12}.  It is likely that the narrow components are contaminated by \Ha\/ from a stronger precursor than is predicted by \citep{morlino12} or by some other background.  \citet{guest22a} fit the X-ray spectrum of the region selected as shocked circumstellar medium on the basis of X-ray line ratios, but that region may be dominated by slower shocks in regions far from the positions of the MUSE spectra.  The proton temperatures from the \Ha\/ profiles and the values of n$_e$t from the X-ray fits would give T$_e$ from Coulomb collisions alone of 1.6 keV, which is well above the value of 0.9 keV from the fit.  The average proper motion expansion speed is 4170 \kms\/ \citep{williams22}, and that speed combined with T$_{e,x}$ = 0.9 keV from \citet{guest22a} yields \teti\/= 0.03.  Because T$_{Coul}$ is almost twice the electron temperature measured from the X-ray spectrum, and because slower shocks probably contribute to the X-ray emission averaged over much of the SNR and reduce the temperature determined from the X-ray spectra, we do not include N103B in the tables or Figure 3, but it is consistent with the low values of \teti\/ in other fast shocks.

\section{1E102.2-7219}   

There are no \Ha\/ profiles for this SNR, apparently because the shock is passing through ionized material.  \citet{xi19} measured the expansion proper motion of 1E102.2-7219, and they fit the X-ray spectra of the shocked ISM in 5 sectors covering over half the circumference of the SNR.  They found temperatures in a narrow range between 0.61 and 0.75 keV and values of n$_e$t of $1.1 -  2.3\times 10^{11}~\rm cm^{-3}~s$.  For a shock speed of 1610 \kms\/ and n$_e$t = $1.7 \times 10^{11} \rm ~cm^{-3}~s$, Coulomb collisions alone could heat the electrons to 1.7 keV, which is above the observed electron temperature. The value of n$_e$t is not unreasonable given an estimated age of 1740 years \citep{banovetz21}, but it is still the least reliably determined parameter in this comparison.

\vspace{5mm}



\bibliography{equilib}{}
\bibliographystyle{aasjournal}

\end{document}